\documentclass{article}
\usepackage{graphicx}
\begin{document}
\begin{large}
\baselineskip 0.26in

\titlepage
\vspace*{0.4in}

\begin{center}
\begin{LARGE}
{\bf An equation-free approach to coupled oscillator dynamics: 
the Kuramoto model example}
\end{LARGE}
\vspace{0.20in}
\end{center}

\begin{center}

{\bf Sung Joon Moon}\\
\vspace{0.1in} Department of Chemical Engineering, 
Program in Applied and Computational Mathematics,
Princeton University, Princeton, NJ 08544\\
Electronic mail address: moon@arnold.princeton.edu\\

\vspace{0.4in}

{\bf Ioannis G. Kevrekidis\footnote{Author to whom correspondence should be directed}}\\
\vspace{0.1in} Department of Chemical Engineering,
Program in Applied and Computational Mathematics; also Mathematics,
Princeton University, Princeton, NJ 08544\\
Electronic mail address: yannis@princeton.edu \\

\vspace{0.4in}

\date{}
\vspace{0.4in}

\begin{abstract}
We present an equation-free multi-scale approach to the
computational study of the collective dynamics of the Kuramoto
model~[{\it Chemical Oscillations, Waves, and Turbulence},
Springer-Verlag (1984)],
a prototype model for coupled oscillator populations.
Our study takes place in a reduced phase space of
coarse-grained ``observables" of the system: the first
few moments of the oscillator phase angle distribution.
We circumvent the derivation of explicit dynamical equations
(approximately) governing the evolution of these coarse-grained 
macroscopic variables; instead we use the {\it equation-free
framework}~[Kevrekidis {\it et al.}, {\it Comm. Math. Sci.}
{\bf 1}(4), 715 (2003)] to computationally solve these
equations without obtaining them in closed form.
In this approach, the numerical tasks for the conceptually
existing but unavailable coarse-grained equations are
implemented through short bursts of appropriately initialized
simulations of the ``fine-scale", detailed coupled oscillator model.
Coarse projective integration and coarse fixed point computations
are illustrated.
\end{abstract}
\end{center}

\newpage

\section{Introduction}

Coupled nonlinear oscillators can exhibit spontaneous emergence
of order, a fundamental qualitative feature of many complex
dynamical systems~[Manrubia {\it et al.}, 2004].
The collective, coarse-grained dynamics of coupled oscillator
populations can provide insights in synchronization
phenomena in many biological, chemical, and physical
systems~[Winfree, 1967; Walker, 1969; Buck, 1988; Gray {\it et al.}, 1989;
Ertl, 1991; Wiesenfeld {\it et al.}, 1996; N{\'e}da {\it et al.}, 2000;
Oliva \& Strogatz, 2001; Kiss {\it et al.}, 2002].
In certain prototype models some exact results for the
collective dynamics, mainly for an order parameter at the
asymptotic states, can be obtained for either a {\em few} or
an {\em infinite} number of oscillators (the so-called continuum
limit)~[Kuramoto, 1975; Kuramoto, 1984; Ariaratnam \& Strogatz, 2001];
for instance, the continuum-limit Kuramoto model is well known
to exhibit a phase transition at a critical coupling strength.
However, general coupled oscillator models are seldom amenable
to mathematical analysis, even for simple coupling topologies,
and the understanding of the coarse-grained (emergent or macroscopic)
dynamics of a given oscillator population remains limited.
Even fundamental questions, such as global stability, spectral
properties of the stationary solution, and finite-number
fluctuations, often remain unanswered~[Strogatz, 2000].

Understanding the coarse-grained behavior of dynamical systems
consisting of a large number of mutually interacting
entities is one of the main goals of statistical mechanics.
In some cases, a fine-scale (or microscopic) model with a large
number of degrees of freedom can be systematically reduced to a
low-dimensional, coarse-grained model describing the dynamics
by a small number of macroscopic variables (observables).
However, such a reduction is rarely available in contemporary
science or engineering dealing with specific complex systems.
One often has to deal with a situation where a reliable
model is available only at a fine-scale level (atomistic,
individual-based), yet he/she is practically interested in the
emergent, system level, coarse-grained dynamics.
Even if a useful coarse-grained model is believed to exist
at a practical closure level, explicit formulas for this
model are often unavailable; the systematic derivation of
closed formulas for such models is difficult without
introducing many simplifying assumptions.
In some cases,
it is also possible that the description of a system simply
does not close at a particular level of coarse-graining.
In the case that a coarse-grained description exists, yet it is
unavailable in closed form, the recently developed {\em equation-free}
framework provides a systematic, computer-assisted approach to
exploring coarse-grained macroscopic behavior without explicit knowledge
of the coarse-grained equations themselves~[Theodoropoulos {\it et al.}, 2000;
Gear {\it et al.}, 2002; Kevrekidis {\it et al.}, 2003;
Kevrekidis {\it et al.}, 2004].
This approach has been successfully combined with several classes of
microscopic models, such as lattice-Boltzmann, kinetic Monte Carlo, and
molecular dynamics simulators, to study the coarse-grained
behavior of heterogeneous catalytic reactions, liquid
crystal rheology, peptide fragment folding, etc.~[Theodoropoulos {\it et al.}, 2000;
Gear {\it et al.}, 2002; Makeev {\it et al.}, 2002;
Siettos {\it et al.}, 2003, Hummer \& Kevrekidis, 2003].
In this paper we illustrate the equation-free approach to
the computational exploration of the coarse-grained dynamics of
the Kuramoto model at the strong coupling limit.

The rest of the paper is organized as follows:
The model under consideration and the regime in which we study it
are described in Sec.~\ref{model}.
Some observations about the coarse-grained behavior and
the equation-free analysis of phase-locked oscillator dynamics is
presented in Sec.~\ref{results}.
Conceptually, the generalization of this approach for different oscillator
models is straightforward; possible shortcomings and limitations will be
addressed in Sec.~\ref{conclusion}, along with future research directions.

\section{Background; the Kuramoto model}
 \label{model}

The Kuramoto model~[Kuramoto, 1975] consists of $N$
equally weighted, all-to-all, phase-coupled limit-cycle oscillators,
where each oscillator has its own natural frequency $\omega_i$
drawn from a prescribed distribution function $g(\omega)$.
We choose $g(\omega)$ to be a Gaussian distribution of standard
deviation 0.1;
however, our analysis is not limited to this particular choice.
The microscopic individual level dynamics is easily visualized
by imagining oscillators as points running around on the unit circle.
Due to rotational symmetry, the average frequency
$\Omega = \sum_{i = 1}^N\omega_i/N$ can be set to 0 without
loss of generality; this corresponds to
observing dynamics in the co-rotating frame at frequency $\Omega$.
The governing equation for the $i$th oscillator phase angle
$\theta_i$ is given by
\begin{equation}
\label{coupledODE}
{d\theta_i\over dt} = \omega_i+{K\over N}\sum_{j = 1}^{N}\sin(\theta_j-\theta_i),~~~ 1 \leq i \leq N,
\end{equation}
where $K \geq 0$ is the coupling strength; in this paper we
arbitrarily set $N = 300$ unless stated otherwise.

It is known that as $K$ is increased from 0 above some critical value
$K_c$, more and more oscillators start to get synchronized (or
phase-locked) until all the oscillators get fully synchronized
at another critical value of $K_{tp}$ (Fig.~\ref{schematic}).
In our choice of $\Omega = 0$, the fully synchronized state
corresponds to an exact steady state of the ``detailed'', fine-scale
problem in the co-rotating frame.
Such synchronization dynamics can be conveniently summarized
by considering the fraction of the synchronized (phase-locked)
oscillators as in Fig.~\ref{schematic}, and conventionally
described by a complex order parameter~[Kuramoto, 1975]
\begin{equation}
re^{i\psi} = {1 \over N}\sum_{j=1}^{N}e^{i\theta_j},
\end{equation}
where the radius $r$ measures the phase coherence, and $\psi$
is the average phase angle.
Some exact results on the synchronization dynamics in the
continuum-limit, in terms of this order parameter, are
available~[Kuramoto, 1984; Strogatz, 2000].\\

{\bf The transition from full to partial synchronization:} 
We begin by restating certain facts about the nature of the 
second transition
mentioned above, a transition between the full and the partial
synchronization regime at $K = K_{tp}$, in the direction of
decreasing $K$.
A fully synchronized state (a ``detailed'' steady state)
in the continuum limit corresponds to the solution to the
following equations, the mean field
type alternate form of Eq.~(\ref{coupledODE})~[Kuramoto, 1975]:
\begin{equation}
\label{mean_field}
{d\theta_i\over dt} = \Omega = \omega_i+rK\sin(\psi-\theta_i),
~~~1 \leq i \leq N,
\end{equation}
where $\Omega$ is the common angular velocity of the fully
synchronized oscillators (which is set to 0 in our case).
It can be rewritten as
\begin{equation}
\label{Eqn3}
{\Omega -\omega_i \over rK}  = \sin(\psi-\theta_i),~~~1 \leq i \leq N,
\end{equation}
where the absolute value of the right hand side (RHS) is
bounded by unity.
As $K$ approaches $K_{tp}$ from above, the LHS for the ``extreme" oscillator
(the oscillator in a particular family that has the maximum value of
$|\Omega -\omega_i|$) first exceeds unity,
and a real-valued solution to Eq.~(\ref{Eqn3}) ceases to exist.
We obtain the detailed steady state solution to Eq.~(\ref{coupledODE})
as a function of $K$ for 300 oscillators by using AUTO2000 (see
Ref.~\cite{auto}), which is based on fixed point computation and 
pseudo-arclength continuation.
The bifurcation diagram plotted in terms of the extreme oscillator 
phase angle reveals that
the transition to the partially synchronized state corresponds to
a turning point bifurcation, of the so-called
saddle node infinite period, or ``sniper" type (Fig.~\ref{sniper}).
Ermentrout and Kopell~[Ermentrout and Kopell, 1984] have studied the
transitions in the partial synchronization regime of a small number
of oscillators, characterizing them as jumps in ``frequency plateaus".

Different random draws of $\omega_i$'s from $g(\omega)$ for a
finite number of oscillators result in slightly different values
of $K_{tp}$.
We observe that $K_{tp}$ appears to follow a Gumbel type extreme
distribution function~[Kotz \& Nadarajah, 2000], just as the maximum
values of $|\Omega -\omega_i|$ do:
\begin{equation}
p(K_{tp}) = \sigma^{-1}e^{-(K_{tp}-\mu)/\sigma}\exp[-e^{-(K_{tp}-\mu)/\sigma}],
\end{equation}
where $\sigma$ and $\mu$ are parameters.

\section{Coarse-grained dynamics of
distribution moments}
 \label{results}

\subsection{Observation}

Our goal is to study the coarse-grained dynamics of an oscillator
population, including the transient dynamics, in terms of coarse-grained
observables.
Since we are dealing with distributions, we naturally consider
the first few moments as candidate coarse-grained variables
(i.e. a kinetic theory-like description).
The moments of a distribution are often easily accessible
in practice, and have a clear physical meaning.
We consider the dynamics of the moments of the phase angle
distribution function $f(\theta)$,
\begin{equation}
\label{moment}
{\cal M}_n \equiv
\left< \left( \theta - \left< \theta \right>\right)^n\right>
= {1\over N}\sum_{j=1}^{N}\left(\theta_j-{1\over N}\sum_{i=1}^{N}\theta_i\right)^n,
\end{equation}
where $n$ is a positive integer, and $\left<~\right>$ means an
ensemble average.
Since we choose to work with symmetric distributions, all the odd
order moments are negligibly small.
To obtain insights on the dynamics in this phase space,
initial configurations of phase angles consistent with four
sets of different values of the first two nonvanishing moments
(${\cal M}_2$ and ${\cal M}_4$) are generated.
These initial phase angle configurations do not have any statistical
correlations with the natural frequencies of the oscillators to which
they are assigned; this is an important issue to which we will
return below.
Evolution of such configurations, obtained by direct numerical
integration of Eq.~(\ref{coupledODE}) reveals that the approach
to a steady state consists of two stages.
During the initial, relatively fast stage both moments decrease, and during a later
slow stage the trajectory gradually approaches to an ultimate steady
state along a common path, essentially a ``slow manifold" of this
dynamical system (Fig.~\ref{approaches}).
The phase angle distributions after the initial stage, when the one-dimensional
slow manifold finally has been approached, are very close
to Gaussian distributions (marked by the dotted line in Fig.~\ref{approaches}).
Once the trajectory approaches this slow manifold, the dynamics can
be effectively described by (or observed on) any one even order moment.

In the following (Sec.~\ref{two_methods}), we will investigate the
dynamics in the moment phase space following the equation-free
approach~[Theodoropoulos {\it et al.}, 2000; Kevrekidis {\it et al.}, 2003;
Kevrekidis {\it et al.}, 2004];
this should be contrasted with the conventional approach of first
attempting to derive closed governing equations for the leading
moments, and subsequently analyzing these equations.

\subsection{Equation-free, coarse-grained multi-scale approach}

The basic idea of the equation-free approach is to utilize
fine-scale simulation {\it as an experiment}, in the form of
a time-stepper: short bursts of appropriately initialized
fine-scale simulations are used to estimate quantities relevant for
scientific computation with the coarse-grained model.
Function evaluations or derivative evaluations for the coarse variables
involved in numerical tasks are thus substituted with estimation
of these quantities, based on appropriately designed computational
experiments with the detailed model.
As a result, traditional continuum numerical techniques are directly 
``wrapped" around
the fine-scale simulator.
The essential steps can be summarized as follows:

\begin{itemize}
\item [1.]
Identify a set of coarse-grained variables (observables) that sufficiently
describe the system dynamics. We expect that a practically useful
model exists {\it and closes} at the level of these coarse observables.
In our analysis, we choose them to be the first two nonvanishing
moments of the phase angle distributions, ${\cal M}_2$ and ${\cal M}_4$.
We will denote the microscopic description by
{\bf u} ($\omega$'s and $\theta$'s in our case) and the macroscopic
description by {\bf U} (${\cal M}_i$'s).
\item [2.] Choose an appropriate $lifting$ operator, $\mu$,
which maps the macroscopic description {\bf U} to one or
more consistent microscopic descriptions {\bf u};
note that this mapping is not uniquely defined.
This operator constructs one or more fine-scale realizations
of the problem consistent with the prescribed values of the 
coarse observables {\bf U} ({\bf u} = $\mu${\bf U}).
\item [3.] Using a lifted, fine-scale realization as a new initial
condition ${\bf u}(t_0)$, run the microscopic simulator to obtain
the values {\bf u}($t_0+\Delta T$) at a later time ($\Delta T > 0$).
This procedure may be repeated for many different realizations
${\bf u}_i(t_0)$ consistent with the same macroscopic initial
condition ${\bf U}(t_0)$, for purposes of ensemble averaging and
variance reduction.
\item [4.] Use an appropriate {\it restriction} operator $M$
which maps the microscopic description to the macroscopic description
${\bf U}(t) = M\left({\bf u}\left(t\right)\right)$.
In our problem the restriction 
operator consists of simple evaluation of (ensemble averages of)
the first two nonvanishing moments of the phase angle distribution,
following their definitions in Eq.~(\ref{moment}).
\end{itemize}

The above procedure results in {\it the coarse timestepper}
$\Phi_{\Delta T}$:
\begin{equation}
{\bf U}(t_0+\Delta T) \equiv \Phi_{\Delta T}({\bf U}(t_0))
= M({\bf u}(t_0+\Delta T)),
\end{equation}
a map from coarse variables at time $t = t_0$ to coarse variables
at a later time $t = t_0 + \Delta T$, constructed through the
microscopic simulation. 
Given such information, obtained for a number of appropriately
chosen coarse initial conditions, it is possible to
implement many scientific computing algorithms to solve the
unavailable coarse-grained equations;
see Sec.~\ref{two_methods} for details of such computations using
two different methods.
An important element of these algorithms is the fact that they
function as protocols for the {\it design of experiments} with the
fine-scale simulator: in the process of their execution they
suggest new computational experiments with the coarse timestepper,
and ``bootstrap" their output toward the ultimate
desired result.
The success of this approach relies (as most of traditional
continuum numerical analysis) on smoothness of the dynamics 
with respect to the coarse variables of interest (in time,
in space and in phase space) and on the identification of proper
lifting and restriction operators.
An extensive discussion of this method can be found in
[Kevrekidis {\it et al.}, 2003].

During the {\it lifting} step, we create phase angle distributions
conditioned on a finite number of prescribed moments.
This is accomplished as follows:
Given $g(\omega)$, phase angles consistent with different values
of the moments are generated by drawing random values from the
inverse cumulative distribution of the following distribution
function
\begin{equation}
\label{randomdraw}
f_I(\theta) = f_g(\theta^2)\left[1+\sum_{p=2}^{n(>2)} a_pS_p(\theta^2)\right],
\end{equation}
of which residual deviation from the Gaussian is expanded
using Sonine polynomials (associated Laguerre
polynomials)~[Chapman, 1970] as basis functions;
$f_g(\theta^2)$ is a Gaussian distribution function with a
prescribed value of ${\cal M}_2$, and $a_p$ is the coefficient of
the $p$th order Sonine polynomial $S_p$.
The value of ${\cal M}_{2p}$ is determined only by $a_p$,
because the Sonine polynomials are otrhogonal with the Gaussian
as the weighting function:
\begin{equation}
\int_{0}^{\infty} e^{-\theta^2}S_p(\theta^2) S_q(\theta^2) d\theta
= \delta_{pq}{\cal N}_p,
\end{equation}
where $\theta$ is generally a $d$-dimensional vector,
$\delta_{pq}$ is the Kronecker delta, and ${\cal N}_p$ is a
normalization constant.
The first three Sonine polynomials are given by
$S_0(\theta^2) = 1$, $S_1(\theta^2) = {1 \over 2}d - \theta^2$,
and $S_2(\theta^2) = {1 \over 8}d(d+2) - {1 \over 2}(d+2)\theta^2
+ {1 \over 2}\theta^4$,
where $d$ is the dimension, which in our case is unity.
Our lifting operator takes us from the values of
a few lower order moments to the values of an equal 
number of Sonine polynomial coefficients, and then to a
consistent phase angle distribution through $f_I(\theta)$.
We consider only even functions for $f_I(\theta)$, due to the
symmetry of $f(\theta)$ with our choice of $g(\omega)$.

One can clearly see in Fig.~\ref{approaches} that observation of
the full system in the ${\cal M}_2-{\cal M}_4$ phase space gives
trajectories crossing one another; this suggests that a deterministic 
description closing with only these two variables is not possible.
However, {\it once the initial fast evolution stage has run its course},
trajectories evolve on an apparently one-dimensional slow manifold,
which suggests that the {\it long-time} coarse-grained dynamics
could be described in terms of a single observable, say ${\cal M}_2$
(or equivalently any other nonvanishing order moment).
In other words, after sufficient time has elapsed,
all the higher moments become slaved to ${\cal M}_2$.
Considering this observation, a more successful lifting operator may be
constructed from the following two steps, when ${\cal M}_2$ is chosen
to be a coarse variable of interest: (1) Drawing phase angles from
an $f_I(\theta)$ satisfying a desired value ${\cal M}_2 = C_1$,
and (2) numerically integrating Eq.~(\ref{coupledODE}) together
with an algebraic constraint ${\cal M}_2 - C_1 = 0$ until the
trajectory approaches the slow manifold.
After this preparatory computation, the lifting step is complete;
the constraint is then removed,
and the system is allowed to evolve in order to evaluate its coarse
timestepper.
This lifting operation results in an augmented problem (ODEs with
an algebraic constraint) that is described by a system of
differential algebraic equations (DAEs) of index 1.
For our illustrative example, 
it is straightforward to solve this set of DAEs by using the Lagrange
multiplier method and the package DASSL (see Ref.~\cite{dassl}).
Direct integration of DAEs showing the relatively fast direct
approach to the ``mature" phase angle distributions on the slow
manifold can be seen in Fig.~\ref{DAEs}.
Once the constrained trajectory approaches the slow manifold,
the constraint is relaxed, and we can observe its subsequent approach
to the steady state {\em along} the manifold. 
 
\subsection{Equation-free computational results}
 \label{two_methods}

We now compute the steady state values of the moments in the
fully synchronized regime ($K > K_{tp}$) using two different
continuum numerical methods in the equation-free framework:
{\it coarse projective
integration} and {\it coarse fixed point computation}.

We first evolve the system toward the stable steady state value of
${\cal M}_2$ using coarse projective integration
[Gear \& Kevrekidis, 2003a; 2003b].
The main assumption is that a macroscopic equation exists and closes
in terms of ${\cal M}_2$; our observations of the long-term
dynamics are consistent with such an assumption.
Starting from a relatively short time ($\Delta t = 0.8$) of direct integration,
restriction of the results [shown as first group of five dots
in Fig.~\ref{PI} (a)] is used to estimate the coarse variable
time derivative.
Smoothness of the trajectory in time (Taylor series, as they appear
in the simplest numerical integration schemes, such as forward Euler)
is then used to ``project" the value of the observable ${\cal M}_2$ in forward
time.
We then {\it lift} with this estimated value
[shown as the fist dot in the next group in Fig.~\ref{PI} (a)],
and the whole procedure is repeated until the steady state is approached.
During the evolution step, we do solve the full system of ODEs;
but we do not solve it {\it for all time}:
short bursts of detailed simulation are used to ``jump"
(and thus save) time, compared to a full integration [a solid
line in Fig.~\ref{PI} (a)].

As we discussed above, it is only after a relatively long time that the
dynamics appear to ``close" with only one observable ${\cal M}_2$;
this is because a certain evolution time is required before all
higher moments of the distribution become slaved to ${\cal M}_2$.
If we are interested in dynamics occurring over shorter time scales,
we need to work with a larger set of observables.
For example, if we are interested in dynamics over {\it shorter} times
than those required for ${\cal M}_4$ to get slaved to ${\cal M}_2$,
we need to work with a coarse-grained model in terms of both
${\cal M}_2$ and ${\cal M}_4$ as coarse variables.
The underlying premise is that the sequence of moments constitutes a
singularly perturbed hierarchy; as longer times elapse, higher moments
get progressively slaved to lower ones. 
Slow manifolds in moment space (graphs of functions, expressing higher
moments as functions of lower ones) embody the closures that allow us
to reduce the dimensionality of long-term dynamics.
In equation-free computation, short simulation of the system itself is
used to bring the system close to these slow manifolds (i.e., establish
the closure numerically).
At this level of coarse description, projective integration can again
be done by lifting consistently with prescribed values of both
${\cal M}_2$ and ${\cal M}_4$ (two constraints in the lifting DAE step)
[Fig.~\ref{PI} (b)]. 

Depending on the computational cost of projection, lifting,
and restriction, these methods have the potential for 
considerable computational savings in problems characterized
by a large separation of time scales (gaps in their eigenvalue spectrum).
In our case, we used a simple least squares fit algorithm for the
estimation and subsequent
projection in time, which required negligible computational effort.
Restriction also required negligible computation. 
However, each lifting step (integrating DAEs for times long enough
to approach the slow manifold) could be relatively expensive.
Working over shorter time scales (more coarse observables, higher
dimensional slow manifolds) alleviates the duration of this preparatory
lifting step.
Finally, it is interesting to note that, in certain cases, forward integration
can be used to evolve the coarse-grained system {\it backward} in
time, possibly converging to certain types
of saddle unstable states~[Gear \& Kevrekidis, 2004a].

\begin{table}[t]
\begin{center}
\begin{tabular}{|c||c|c|}
\hline
iteration & ${\cal M}_2$ & \mbox{$|{\cal M}_2-{\cal M}_2^{ss}|/{\cal M}_2^{ss}$} \\
\hline\hline
0 & 1.0000$\times 10^{-2}$ & 5.6768$\times 10^{-1}$ \\
1 & 3.0362$\times 10^{-2}$ & 3.1261$\times 10^{-1}$ \\
2 & 2.3179$\times 10^{-2}$ & 2.0751$\times 10^{-3}$ \\
3 & 2.3096$\times 10^{-2}$ & 1.5131$\times 10^{-3}$ \\ 
\hline
\end{tabular}
\caption{\label{table1}
Coarse fixed point computation at $K = 0.7$, using the
Newton-Raphson method for ${\cal M}_2$ only.
The steady state value ${\cal M}_2^{ss}$, the variance of
detailed steady state distribution, is $2.3131\times 10^{-2}$.
The full system, Eq.~(\ref{coupledODE}), was integrated for $\Delta t = 2.0$
at each iteration step, and nearby conditions separated by
$\epsilon = 1.0\times 10^{-5}$ were used for linearization.
}
\end{center}
\end{table}

\begin{table}[t]
\begin{center}
\begin{tabular}{|c||c|c|c|}
\hline
iteration & ${\cal M}_2$ & ${\cal M}_4$ & \mbox{$|{\cal M}_2-{\cal M}_2^{ss}|+|{\cal M}_4-{\cal M}_4^{ss}|$} \\
\hline\hline
0 & 1.8000$\times 10^{-2}$ & 9.0000$\times 10^{-4}$ & 5.8386$\times 10^{-3}$ \\
1 & 2.3314$\times 10^{-2}$ & 1.8803$\times 10^{-3}$ & 4.5518$\times 10^{-4}$ \\
2 & 2.2902$\times 10^{-2}$ & 1.6057$\times 10^{-3}$ & 2.3052$\times 10^{-4}$ \\
\hline
\end{tabular}
\caption{\label{table2}
Coarse fixed point computation at $K = 0.7$, using the
Newton-Raphson method both for ${\cal M}_2$ and ${\cal M}_4$.
The steady state value ${\cal M}_4^{ss}$, obtained from the 
detailed steady state distribution, is $1.6076\times 10^{-3}$.
The full system was integrated for $\Delta t = 2.0$
at each iteration step, and nearby conditions separated by
$\epsilon = 1.0\times 10^{-5}$ for ${\cal M}_2$ ($1.0\times 10^{-6}$
for ${\cal M}_4$) were used for linearization.
}
\end{center}
\end{table}

In addition to coarse projective integration as a means of approaching
the steady state {\it in time}, we also approximate it using
a coarse contraction mapping, based on the Newton-Raphson  
method~[Kevrekidis {\it et al.}, 2003];
we compute the solutions ${\cal M}_i^{ss(CG)}$ to the following
equations (of which explicit forms are unavailable):
\begin{equation}
\Phi_{\Delta T}({\cal M}_2^{ss(CG)}) - {\cal M}_2^{ss(CG)} = 0
\end{equation}
or
\begin{equation}
\Phi_{\Delta T}
\left( \begin{array}{c}
{\cal M}_2^{ss(CG)} \\
{\cal M}_4^{ss(CG)} \end{array} \right) 
-
\left( \begin{array}{c}
{\cal M}_2^{ss(CG)} \\
{\cal M}_4^{ss(CG)} \end{array} \right) 
=0,
\end{equation}
depending on the level of closure being considered,
where the superscript stands for coarse-grained steady state.
In this computation, residual and derivative evaluations required
for each Newton-Raphson iteration are obtained by short bursts of
direct integration of the microscopic model for $\Delta T = 2.0$,
starting at {\it nearby initial conditions}.
Within only few Newton-Raphson iterations, an accurate
{\em solution} to the unavailable equation can be found
(Tables~\ref{table1} and~\ref{table2}).
As a byproduct of this computation, upon convergence, the eigenvalues
of the coarse linearization at steady state can be computed.
In this case, when a steady state of the coarse problem corresponds
to a true steady state of the fine-scale one, the eigenvalues of 
the two are related (one expects to find, from the coarse
procedure, the leading slow eigenvalues of the detailed problem
linearization, discounting the neutral eigenvalue at 0 that
corresponds to rotational symmetry).
As the number of coarse-grained variables increases, estimating each
element of the coarse Jacobian becomes increasingly cumbersome; matrix-free
methods of iterative linear algebra, like Newton-GMRES, are then used to
solve the equations arising in the equation-free fixed point computation.
These methods (such as Newton-GMRES; see the monograph [Kelley, 1995])
are based on the
estimation of matrix-vector products, using short bursts of simulation
with appropriately chosen {\it perturbations} of a given coarse initial
condition (see also the Recursive Projection Method~[Shroff \& Keller, 1993]).

\section{Conclusions}
 \label{conclusion}

The long-term collective dynamic behavior of the Kuramoto model
at the continuum (or thermodynamic) limit is often described
in terms of an order parameter $r$~[Strogatz, 2000].
In this work we chose to observe the (long-time) collective
dynamics for finite assemblies of coupled oscillators
using a different set of coarse-grained variables:
low order moments of the phase angle distribution.
We found that, after sufficient evolution time has elapsed,
the dynamics lie on a slow manifold parametrized by a
few of these moments.
The longer the dynamics have evolved, the lower the dimension
of this manifold becomes, as higher order moments get 
progressively slaved to lower order ones.
As a result, if we are interested in relatively long-term
dynamics, we can work with less macroscopic observables 
in our lifting and restriction schemes.
Two levels of closure were explored (parametrized by only one moment,
and by two moments of the phase angle distribution respectively);
shorter lifting computations were required for the
two-moment description.
We circumvented the derivation of closed governing equations
for these coarse-grained variables, following the equation-free,
coarse-grained multi-scale approach.
Using this approach, we demonstrated coarse projective integration
as well as coarse fixed point computation (followed by coarse
stability analysis) in the fully synchronized regime.

Our approach can, in principle, be extended in a straightforward
manner to explore the dynamics of other coarse-grained variables
of this model, or more complicated coupled oscillator models,
as long as the long-term coarse-grained dynamics exhibit smooth
low-dimensional behavior
(i.e. are characterized by a slow, attracting manifold in the 
corresponding phase space).
A very important step is the identification of appropriate lifting and
restriction operators,
that allow us to (approximately) initialize the fine-scale
system close to the slow manifold, and to compute the evolution
of its macroscopic observables.

As we saw in Fig.~\ref{approaches}, simply prescribing two moments
of the distribution and using Eq.~(\ref{randomdraw}) does not
provide a good enough lifting; temporal trajectories of the coarse
variables cross one another.
We obtained a better lifting by introducing an additional preparatory
step: evolution of the system constrained on the observables,
until it approached the low-dimensional, slow manifold.
This required the integration of a system of DAEs (in the
spirit of algorithms like SHAKE~[Ryckaert {\it et al.}, 1977]  and
umbrella sampling~[Torrie \& Valleau, 1974]).
For this problem it was easy to augment the fine-scale model
to introduce such a constraint; algorithms guiding to a point
on the manifold {\it without} the imposition of an explicit
constraint have also been devised, which was used for the detailed
simulator as an unmodified ``legacy code"~[Gear \& Kevrekidis, 2004b;
Gear {\it et al.}, 2004c].

Suppose that the steady state phase angle distribution is known (e.g.
from a long simulation), and we lift consistently with {\it all} of its
moments, not just the first two.
When, in a simple lifting step, angles drawn from this known distribution
are arbitrarily assigned to oscillators with different natural
frequencies, the fine-scale evolution will move away from the steady
state before eventually returning to it.
Even though moments are good variables for parameterizing the slow
manifold, lifting with them is difficult; one needs to evolve
constrained on the prescribed moments to obtain ``mature" distributions,
that both have these moments {\it and} are close to the 
slow manifold.
Observing the evolution of the state during the constrained integration 
sheds light on the nature of difficulty in this lifting 
problem: we observed that during the initial evolution stages correlations
{\it between the phase angle and the oscillator natural frequency}
develop.
% (see Fig.~\ref{DAEs}).
%
These correlations are destroyed when we randomly  assign angles from
a given distribution to oscillators with different natural frequencies.
This strongly suggests that {\it different coarse-grained
observables} that take these correlations into account must be sought, so that
lifting can be implemented in a simpler and less computationally
complicated/expensive fashion.
We propose that this can be done through the framework of the Wiener's
polynomial chaos, where both the phase angles and the natural
frequencies are treated as random variables, and the former is expanded
in terms of Hermite polynomials of the latter;
the expansion coefficients are chosen to be the coarse variables 
[Moon {\it et al.}, 2005].

In our current work we clearly saw that the selection of good coarse-grained
observables that parametrize the slow manifold, and the construction
of a good lifting operator are vital for the success and
competitiveness of this approach compared to direct simulation.
For the examples studied in this paper, only a small number of
macroscopic observables were enough to answer the coarse-grained
questions of interest.
Equation-free methods like the {\it gap-tooth scheme}
[Gear \& Kevrekidis, 2003c] and {\it patch dynamics}
[Samaey {\it et al.}, 2004] can be useful for problems
whose dynamics are described
in terms of smooth {\it fields} of coarse-grained observables.
The review papers [Kevrekidis {\it et al.}, 2003; 2004] contain detailed
descriptions of equation-free methods, as well as references to an
extensive set of applications where the algorithms have been applied
and experience with their properties and limitations that
has been gained.

{\bf Acknowledgements}.
This work was partially supported by AFOSR (Dynamics and Control)
and an NSF ITR grant.
It is a pleasure to acknowledge discussions with Dr. Dongbin Xiu,
Prof. Roger Ghanem and the long-term collaboration with 
Prof. C. William Gear in the development of equation-free algorithms.

\end{large}

\begin{figure}[t]
\begin{center}
\includegraphics[width=.6\columnwidth]{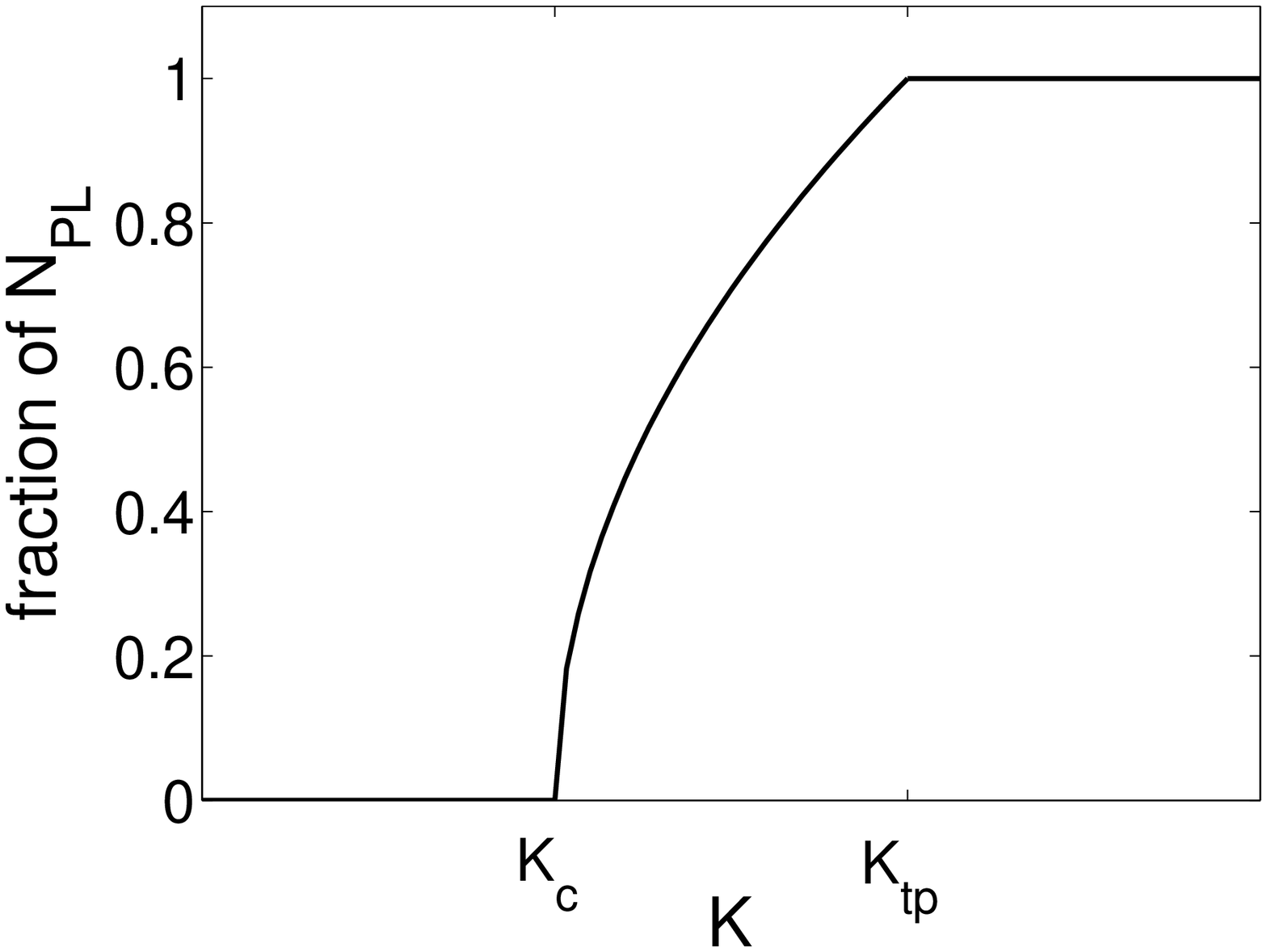}
\caption{
\label{schematic}
Qualitative schematic of the fraction of phase-locked oscillators
in the Kuramoto model; there exist two critical values in K,
the coupling strength.
$N_{PL}$ is the number of phase-locked oscillators.
}
\end{center}
\end{figure}

\begin{figure}[t]
\begin{center}
\includegraphics[width=.6\columnwidth]{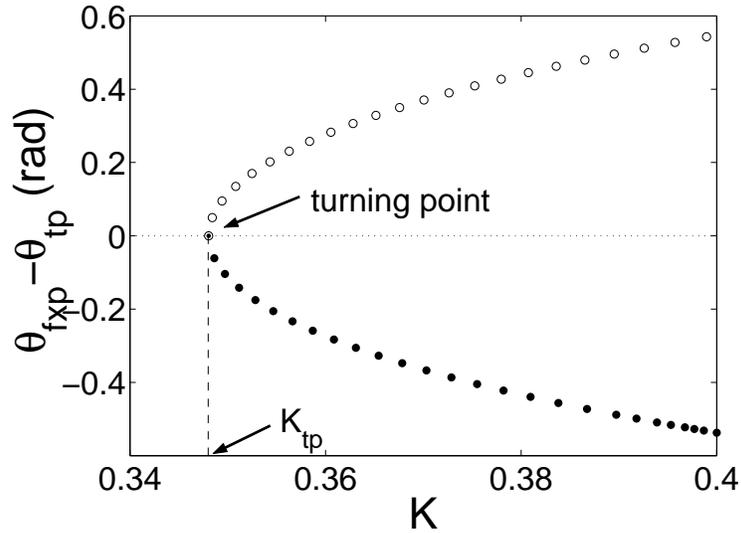}
\caption{
\label{sniper}
Bifurcation diagram of the oscillator that desynchronizes first with
decreasing $K$ (this is the one with the maximum value of
$|\omega_i-\left<\omega_i\right>|$, the `` extreme'' oscillator).
The transition from the full to the partial synchronization regime
at $K = K_{tp}$ corresponds to a turning point of this oscillator
phase angle at steady state ($\theta_{fxp}$),
where $\theta_{tp}$ is the phase angle at the turning point.
For the extreme oscillator, a saddle state and a stable node 
state collide in a saddle-node infinite period (sniper) 
bifurcation at $K = K_{tp}$.
}
\end{center}
\end{figure}

\begin{figure}[t]
\begin{center}
\includegraphics[width=.6\columnwidth]{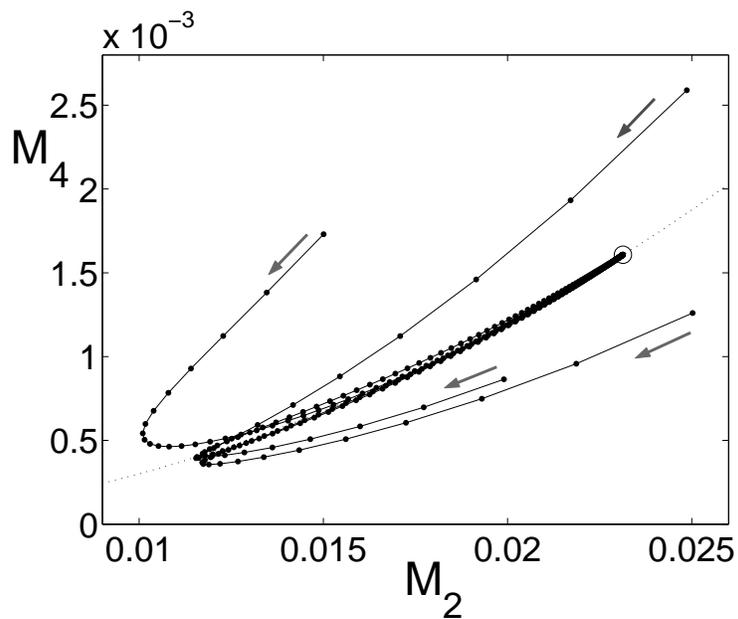}
\caption{
\label{approaches}
Trajectories of 300 oscillators ($K = 0.7$) in the phase space
of ${\cal M}_2$ and ${\cal M}_4$, started from four different
sets of values, but with the same natural frequency distribution.
Each dot is separated by 0.1 in time
(arrows indicate the direction of increasing time).
Initially, $g(\omega)$ and $f(\theta)$ have no correlations.
In all cases, both moments decrease first (during $t < 1$), and
then gradually increase toward a steady state 
[(${\cal M}_2^{ss}, {\cal M}_4^{ss}$) = (2.3131$\times 10^{-2}$,
1.6076$\times 10^{-3}$) marked by a circle].
The latter slow evolution occurs along a ``slow manifold", which nearly 
coincides with the curve for Gaussian distributions (dotted line,
${\cal M}_4 - 3{\cal M}_2^2 = 0$).
}
\end{center}
\end{figure}

\begin{figure}[t]
\begin{center}
\includegraphics[width=.6\columnwidth]{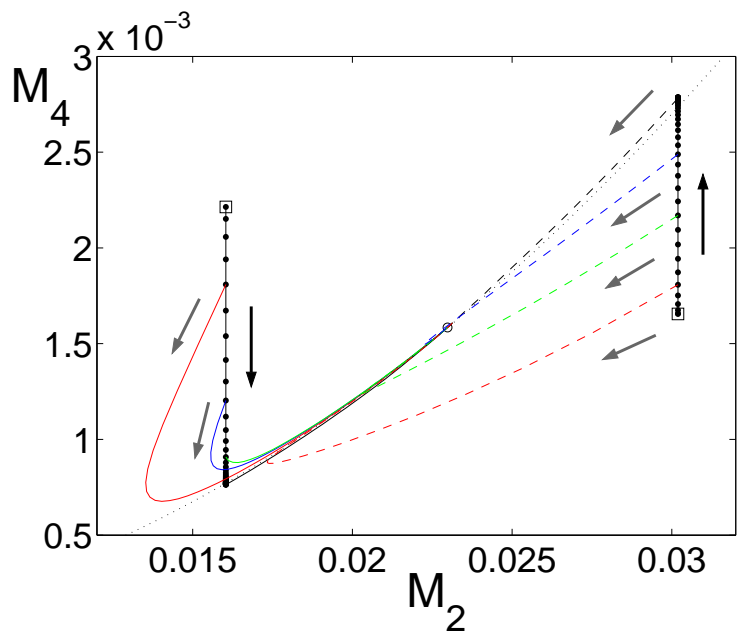}
\caption{
\label{DAEs}
(Color) Two different random initializations of oscillators (marked
by squares) are directly guided (shown by dot-dashed black lines)
to the slow manifold by solving Eq.~(\ref{coupledODE}) along
with a constraint on ${\cal M}_2$ for $K = 0.7$.
Each dot is separated in time by 0.1.
Poor lifting results from relaxing the constraint before the slow
manifold is approached
(colored lines, constraint lifted at various times $t =$ 0.5, 1, 1.5,
and 6 for red, blue, green, and black line, respectively).
Arrows indicate the direction of increasing time.
}
\end{center}
\end{figure}

\begin{figure}[t]
\begin{center}
\includegraphics[width=.6\columnwidth]{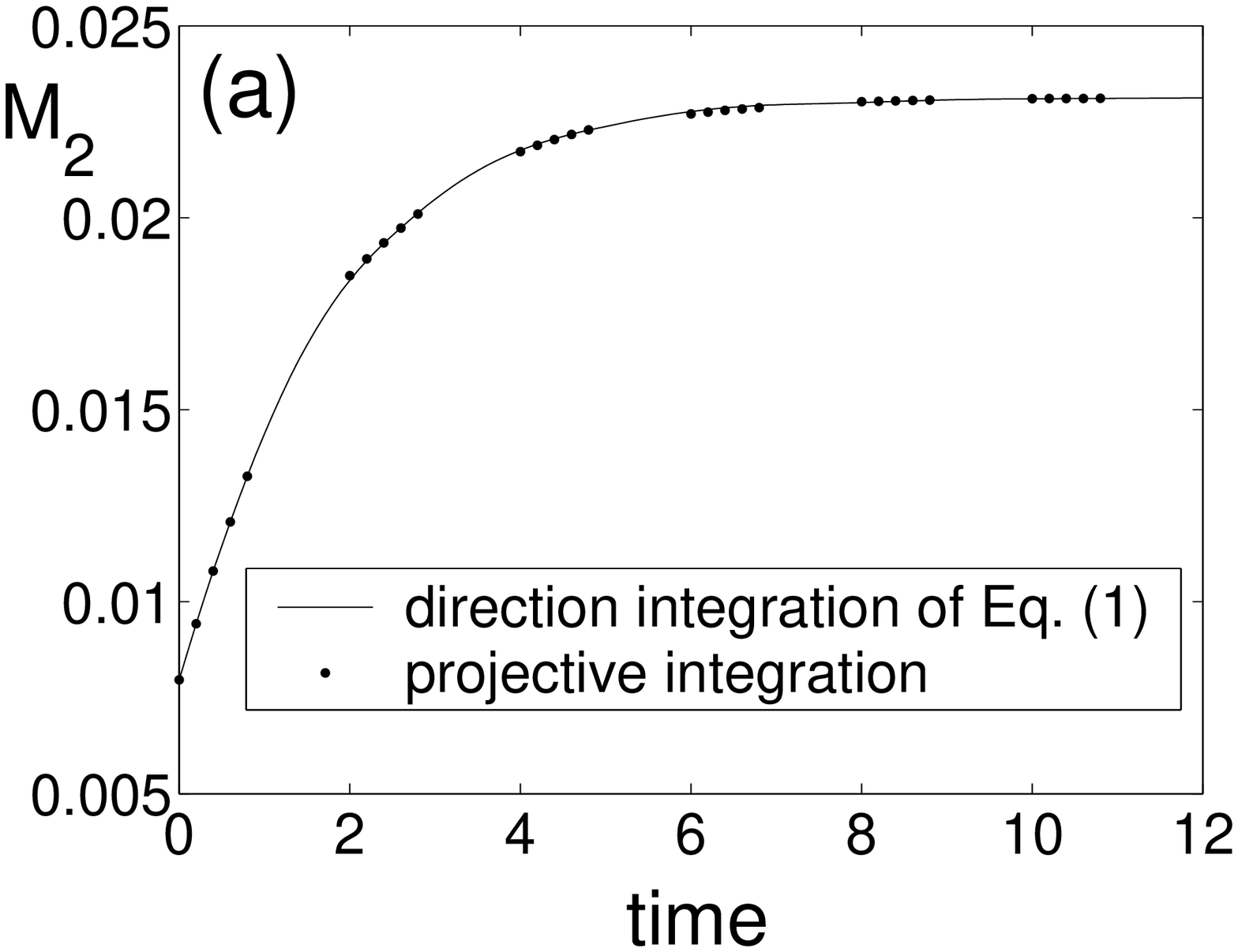}

\includegraphics[width=.58\columnwidth]{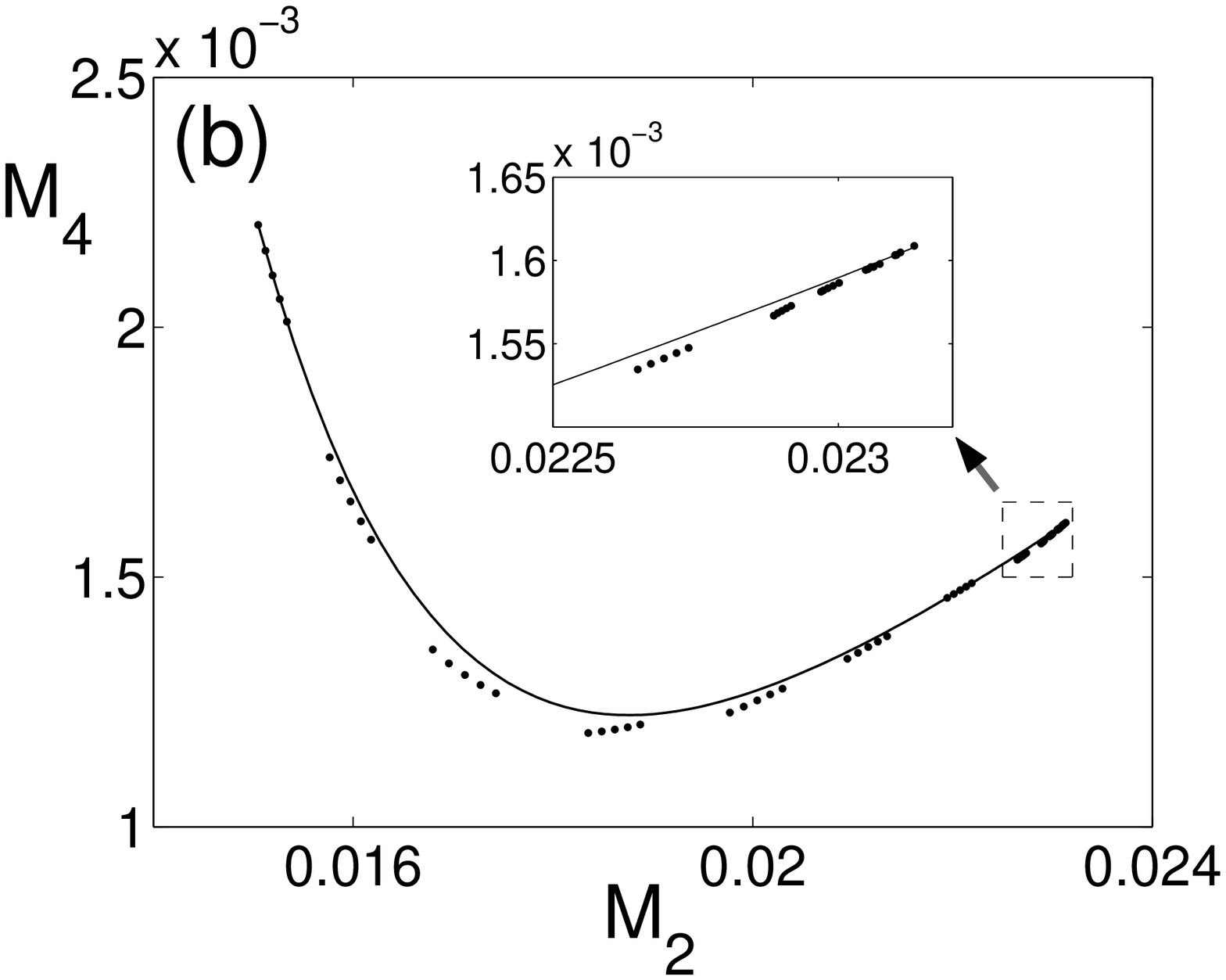}
\caption{
\label{PI}
Transient approach to the steady state ($K = 0.7$) computed
through coarse projective integration (shown as dots), taking
(a) only ${\cal M}_2$, and (b) both ${\cal M}_2$ and ${\cal M}_4$
as coarse variable(s).
Full, fine scale integration is shown for comparison (solid line).
In the projective integration computation, Eq.~(\ref{coupledODE})
is integrated only for the time intervals marked by dots, and
the system is restricted to ${\cal M}_2$ (and ${\cal M}_4$)
at uniformly distributed times (five dots in each group).
The last three points are used to estimate the future value(s);
lifting is then performed again, and the procedure repeated.
For each lifting step, the same $g(\omega)$ was used.
The inset in (b) shows the blow up near the steady state, where
larger projection time steps are used for faster convergence.
}
\end{center}
\end{figure}

\end{document}